\documentclass[9pt,twocolumn,twoside]{opticajnl}

\journal{ol} 

\setboolean{shortarticle}{true}

\usepackage{lineno}

\title{Fabrication method of a low-loss plasmonic waveguide containing both “plasmonic- friendly” and “plasmonic- unfriendly” metals}

\author[1,*]{Vadym Zayets}
\author[2]{Iryna Serdeha}
\author[2]{Valerii Grygoruk}

\affil[1]{Platform Photonics Research Center, National Institute of Advanced Industrial Science and Technology, Tsukuba, Japan}
\affil[2]{Educational and Scientific Institute of High Technologies, Taras Shevchenko National University of Kyiv, Kyiv, Ukraine}

\affil[*]{Corresponding author: v.zayets@gmail.com}

\begin{abstract}
Fabrication technology, which allows a substantial decrease of the plasmonic propagation loss for both  “plasmon- friendly” metals like Au, Cu or Al and “plasmon- unfriendly” metals like Co, Fe or Cr, has been developed and experimentally demonstrated. Optimization of the optical confinement is used to reduce the propagation loss below 1 dB per plasmonic device. 
\end{abstract}

\setboolean{displaycopyright}{true}

\begin{document}

\maketitle

The fabrication of a denser Photonic Integrated Circuit (PIC) allows to increase both the speed and the functionality of data processing optical circuits. In the case of an electronic circuit, the size of MOSFET transistors is very small and billions of transistors can be integrated into one electrical circuit. Because of the dense integration, the electronic devices have a complex functionality at a low cost.  In comparison to the electronic components, the size of the optical components is not as small, and a dense integration of optical components is still challenging.  The plasmonic devices might be the key for a denser integration, because of their small size of about 10-30 $\mu m$.  The uniqueness of a surface plasmon is its topological nature, meaning that the properties of a surface plasmon are defined by the optical properties of a surface, but not of the bulk as in the case of a conventional optical waveguide.  As a result, even a small change of optical properties in close proximity of the interface substantially modifies plasmon`s propagation constants. This makes it possible to modulate light within the propagation distance of only a few wavelengths and, therefore, to fabricate a plasmonic device \cite{Shimizu2018,Zayets2015,Armelles2013,ZIA2006,Sorger2012,Sorger2021,2009AtwaterPlasMod,2013AtwaterPlasMod} of a substantially smaller size than a device, which is made from conventional optical waveguides. 

Another reason, why plasmonic devices are important for PIC, is that a plasmonic device is very fast. An operation speed of 170 GHz of a plasmonic modulator has already been demonstrated \cite{Hoessbacher2017} and is still very far from the speed limit of a plasmonic device \cite{Amin2020,Hoessbacher2017,Haffner2018}. Plasmonic devices are very fast because their length is short and they do not have the problem of phase mismatch. Additionally, the capacitance of the controlling electrodes is small, because of the small size of a plasmonic device.

The optical loss is a key issue for any plasmonic device and should be carefully addressed. A metal absorbs light and, therefore, the propagation loss is unavoidable for a surface plasmon. However, when the design and fabrication technology of a plasmonic device are optimized, the loss due to the light absorption by the metal is negligibly small  and does not degrade the plasmonic device \cite{Shimizu2018,Zayets2015,ZayetsSPIE2022}. Since  light is absorbed by the metal in a plasmonic device, the choice of the metal seems to be very important and it might be suggested that only a few metals can be used in a low-loss plasmonic structure. This Letter clarifies this misconception and describes the technological method of fabricating a plasmonic device with a negligibly small propagation loss, in which a variety of different metals can be used.  The successful experimental demonstration of the proposed method of the loss reduction in a plasmonic device containing "plasmonic-unfriendly" cobalt is described. All calculations below are done by the method of rigorous solution of Maxwell equations in a multilayer structure \cite{Zayets2012tMO,Zayets2015,Zayets2012Isolator}, in which no approximations or simplifications have been used. 

\begin{figure}[h]
\begin{center}
\includegraphics[width=8.5cm]{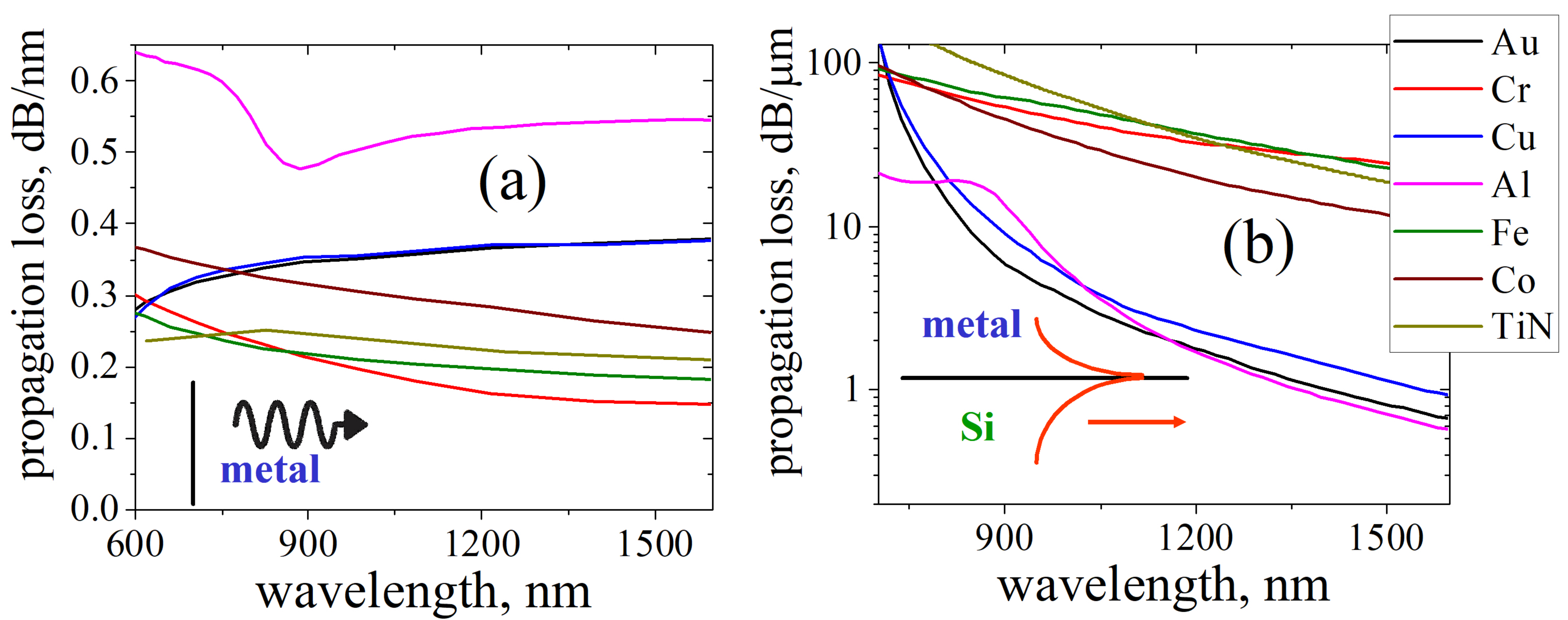}
\end{center}
\caption{\label{fig:fig1} 
(a) Propagation loss for light propagating in the bulk of a metal (b) Propagation loss of a surface plasmon propagating along a $metal/Si$ interface. Note, the loss unit in each graph is different due a substantially smaller loss of a plasmon in comparison to bulk light propagation.
}
\end{figure}



At present, gold is the first choice as a metal material for the plasmonic device and most present plasmonic devices are made of gold \cite{Armelles2013,Kornienko2020,Merzlikin2016,Amin2020,Hoessbacher2017,Haffner2018}. The gold has a yellowish color, in fact, due to plasmonic absorption. It would appear to be  the best choice for a plasmonic device. However, gold is not a good material for fabrication technology. It has a poor adhesion with other material and a large diffusion coefficient. It is important to investigate whether other metals can be used for a plasmonic device instead of gold. 

There is a misconception that there are “plasmon- friendly” metals like $Au$, $Al$, $Cu$, in which the absorption of light is weakest and only which should be chosen for use in plasmonic devices, and  there are “plasmon- unfriendly” metals  like $Co$, $Fe$, $Cr$, in which light absorption is much stronger and which should not be used in a plasmonic device. Figure \ref{fig:fig1}(a) shows the calculated propagation loss of light propagating in the bulk of a metal using optical constants measured in Ref.\cite{Johnson1974}. In fact, the calculations show that the absorption loss in the “plasmon- friendly” metals is larger than in the “plasmon- unfriendly” metals. The smallest absorption is in the “plasmon- unfriendly” chromium and the largest absorption is in the “plasmon- friendly” aluminum.

Figure \ref{fig:fig1} (b) shows the calculated propagation loss of a surface plasmon propagating along a metal/ Si interface. Indeed,  the propagation loss is moderate  (about 1 $dB/\mu m$ at $\lambda$=1550 nm) in the plasmonic structure containing either $Au$, $Al$ or  $Cu$.  In the case of a 5-$\mu m$- long plasmonic device made of these metals, the insertion loss due to metal absorption is about 5 dB. It is still large but could be acceptable for some applications. In contrast, the propagation loss is extremely large (~20 $dB/\mu m$) in the plasmonic structure containing either $Fe$, $Co$ or $Cr$. Even in a 1-$\mu m$-long plasmonic device, nearly all of the light is absorbed and, as a result, such plasmonic structure cannot be used in a PIC. 

 The data shown in Fig. \ref{fig:fig1}(b) may be incorrectly interpreted to indicate that it is true that there are “plasmonic-friendly” and “plasmonic-unfriendly” metals, because  the  plasmonic loss  is substantially smaller in $Au$, $Al$ and $Cu$ than in $Co$, $Fe$ and $Cr$.  In order to clarify  this issue it  is necessary to understand the reason why the absorption of a plasmon and the light absorption in the bulk of metal are so different and the reason why the absorption  of a plasmon in $Al$ is smallest (Fig.\ref{fig:fig1}(b), even though  the light absorption in the bulk of  $Al$  is  largest (Fig. \ref{fig:fig1}(a)).

\begin{figure}[t]
\begin{center}
\includegraphics[width=8.5cm]{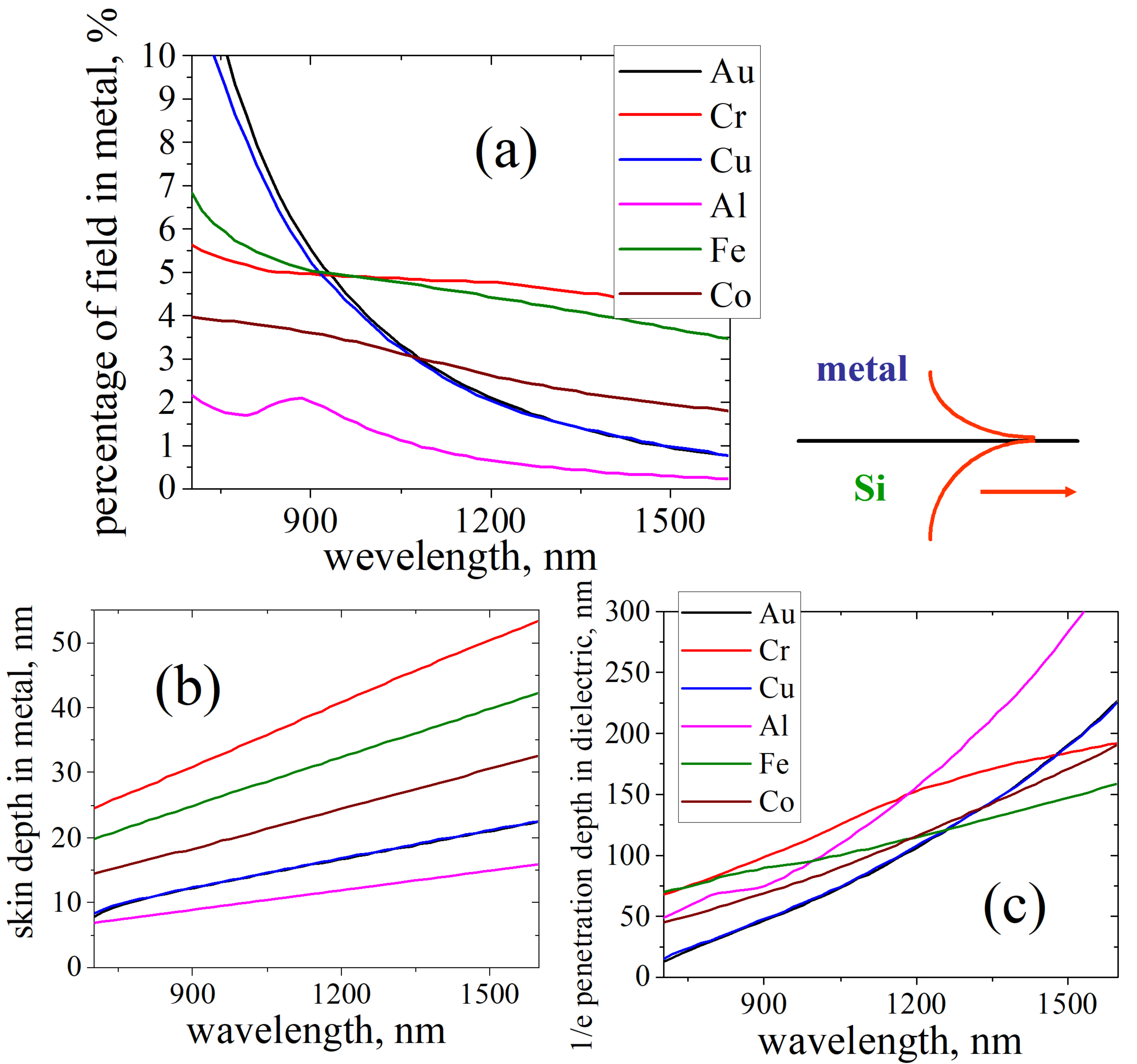}
\end{center}
\caption{\label{fig:fig2} 
Surface plasmon at a $metal/Si$ interface. (a) Percentage of optical field of the plasmon, which is inside of the metal; (b) skin depth in the metal; (c) 1/e penetration depth in $Si$.
}
\end{figure}

Figure \ref{fig:fig2}(a) shows the relative amount of optical field in metal. In the case of iron and chromium, the amount of optical field inside a metal is large reaching 5 \% . In contrast, the amount is less than 0.5 \% in the case of aluminum. That is the reason why the plasmonic absorption is substantially smaller in aluminum in comparison to the iron and chromium despite the bulk absorption in Al being substantially larger.

It is not only the bulk absorption in a metal, but also the refractive index of the metal that determines the absorption of a surface plasmon. Figures \ref{fig:fig2}(b) and \ref{fig:fig2}(c) show the skin depth in the metal and the penetration depth of the optical field into the dielectric. Both the smaller skin depth in the metal and the larger penetration depth into Si are reasons why the relative amount of light penetrating in Al is much smaller than in $Fe$ and $Cr$.

The example of $Al$, in which  the plasmonic  loss is smallest despite of its largest  bulk absorption, indicates a possibility of engineering of a plasmonic structure for a smaller loss by optimizing the field distribution of a surface plasmon and, specifically, by the reduction of the amount of light field  penetrating  into the metal.

\begin{figure}[h]
\begin{center}
\includegraphics[width=8.2cm]{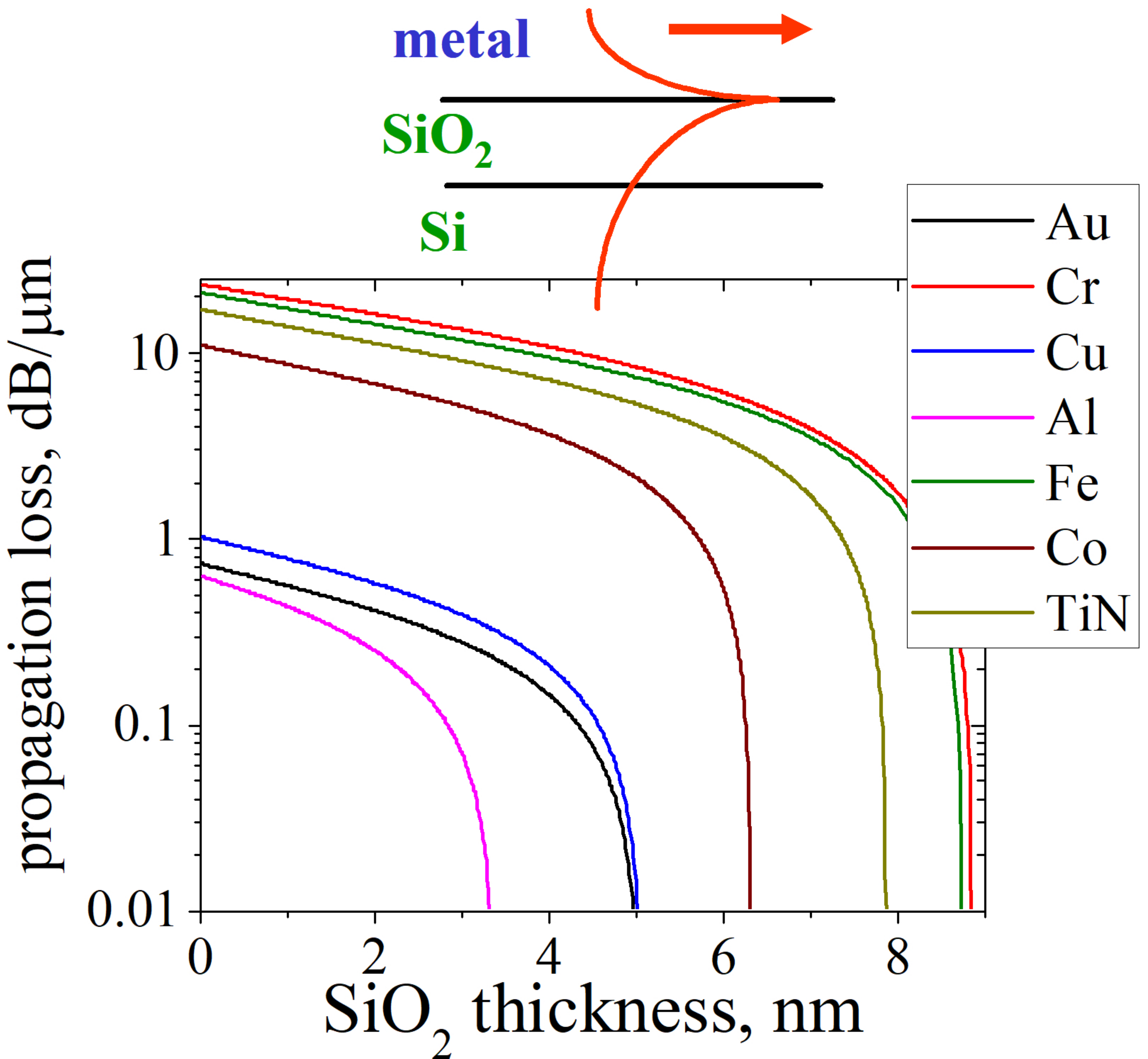}
\end{center}
\caption{\label{fig:fig3} 
Propagation loss of a surface plasmon propagating along a $metal/SiO_2/Si$ interface. For each metal there is an optimum $SiO_2$ thickness, at which the plasmonic loss becomes small.
}
\end{figure}

The amount of light penetrating into the metal can be engineered and reduced in a more complex plasmonic structure, which contains more dielectric layers. For example, when a very thin $SiO_2$ layer is inserted at the $Si/metal$ interface. Figure \ref{fig:fig3} shows the calculated propagation loss in this plasmonic structure vs. $SiO_2$ thickness. For both the “plasmon- friendly” and “plasmon- unfriendly” metals, the propagation loss can be reduced down to 0.01 dB/$\mu m$ at an optimum $SiO_2$  thickness.  The optimum $SiO_2$  thickness is different for each metal. Such a small loss means that for a 10-$\mu m$-long plasmonic device the absorption loss is only 0.1 dB, which is negligible in comparison to the unavoidable coupling loss.


The optimum thickness is a little smaller than the cutoff thickness.  The structure, in which the $SiO_2$  insertion layer is thicker, has no plasmon confinement and does not support plasmon propagation along the surface. The incident light is reflected and absorbed by the metal in such a structure. In case of a substantially-thicker $SiO_2$ layer (/~500 nm), there is another cutoff thickness, above which the structure supports plasmon propagation again \cite{Zayets2012Isolator}. 

Figure \ref{fig:fig3} clarifies an important fact that the propagation loss of a surface plasmon can be reduced to a nearly- equal low number for both  the “plasmonic- friendly” and “plasmonic- unfriendly” metals. The intrinsic absorption of light in the bulk of the metal is only one of several parameters, which define the propagation loss of a surface plasmon.

\begin{figure}[t]
\begin{center}
\includegraphics[width=8.5cm]{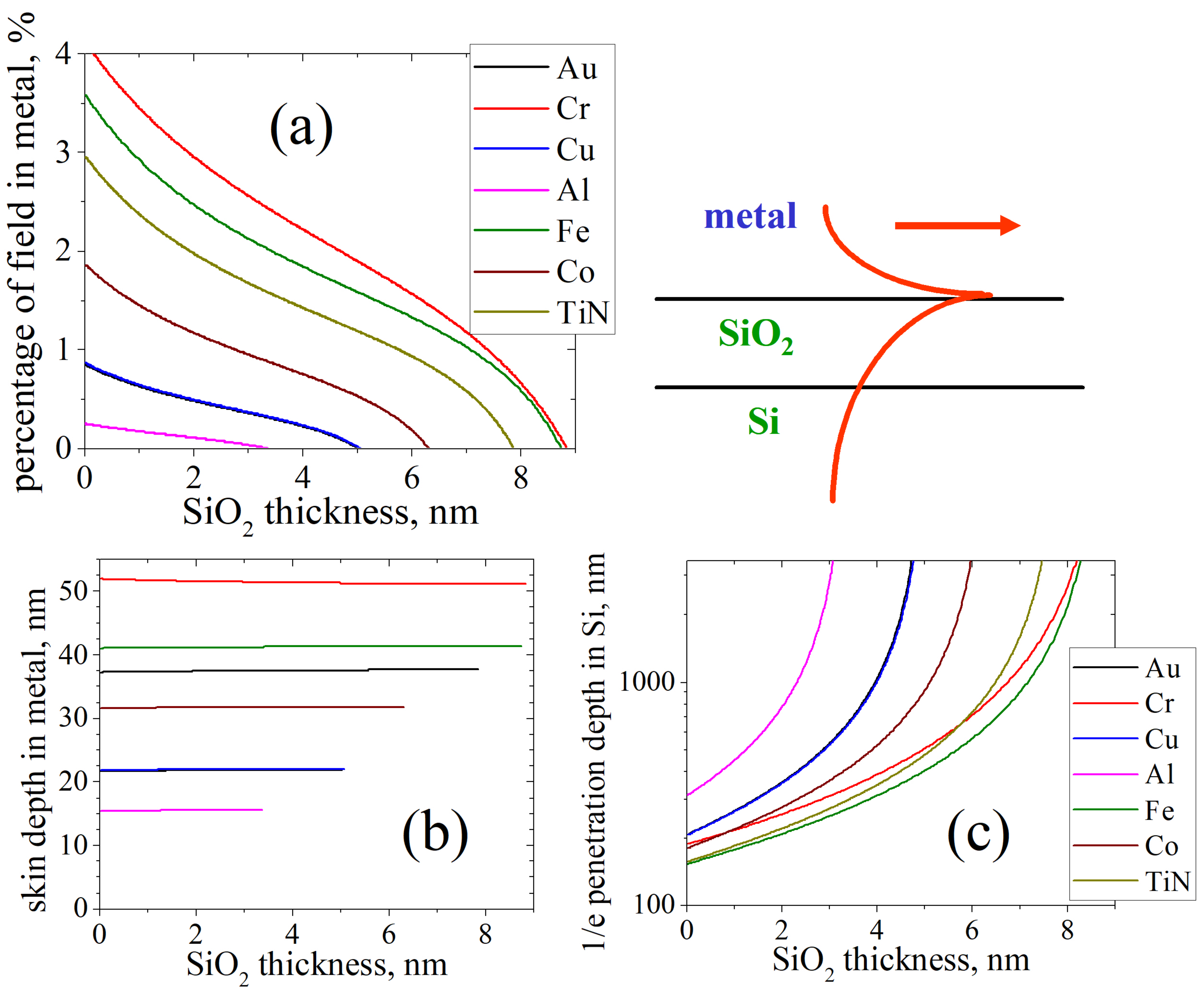}
\end{center}
\caption{\label{fig:fig4} 
Surface plasmon at a $metal/SiO_2/Si$ interface. (a) Percentage of optical field of the plasmon, which is inside of the metal; (b) skin depth in the metal; (c) 1/e penetration depth in Si. $\lambda$ =1550 nm.
}
\end{figure}


Figure \ref{fig:fig4}(a) shows the percentage of light penetrating into the metal as a function of the $SiO_2$ thickness. At the optimum thickness, the penetration is the smallest. Figures \ref{fig:fig4}(b) and \ref{fig:fig4}(c) show the skin depth in the metal and the penetration depth in the dielectric. If the skin depth in the metal is practically independent of the $SiO_2$ thickness, the penetration depth in the dielectric increases substantially near the optimum $SiO_2$ thickness. The increase of the penetration depth into the dielectric is the reason why the percentage of optical field in metal decreases with an increase of the $SiO_2$ thickness (Fig.\ref{fig:fig4}(a)).

 This means that there is a trade-off between the volume of a surface plasmon and the plasmon propagation loss, which cannot be overcome in any metal. The smaller the loss is, the larger plasmon volume becomes. The larger plasmon volume is not a problem for many plasmonic applications. For example, for a data- processing application, the short length of a plasmonic device and the large magnitude of the electro-optical or the magneto-optical effect are important. The volume of a surface plasmon is not as important.

Figure \ref{fig:fig3} shows that, at least theoretically, it is possible to reduce the plasmonic propagation loss to an ever- small number. However, practically there are technological limitations, which limit the possible reduction. The first limitation is the fabrication precision and the roughness of the thin inserted layer. Near the region of the smallest loss, the dependence of loss on the $SiO_2$ thickness becomes very steep and a tiny deviation of the $SiO_2$ thickness from the optimum thickness causes a substantial propagation loss.  The required fabrication precision is more severe for $Cr, Fe, Co$ than for $Au, Al, Cu$, because of their steeper slopes in Fig. \ref{fig:fig3}.

Another technological limitation for the loss reduction is the penetration depth of the optical field into the dielectric. It should roughly not exceed about 10 $\mu m$. Otherwise; the plasmonic properties may become dependent on inhomogeneities and imperfections of the substrate. This limits a possible reduction to about 0.01 dB/$\mu m$ at least in the case of the metals, which were used in calculations shown in Fig.\ref{fig:fig3}.


As an example of the effectiveness of the proposed technology, we have fabricated a low-propagation-loss plasmonic waveguide made of a “plasmonic-unfriendly” $Co$ integrated with a $AlGaAs$ waveguide. As an insertion layer, the 17- nm- thick $MgO$ layer was used. The experimentally-measured plasmon' propagation loss was reduced down to 0.17 dB/$\mu m$ \cite{Zayets2015}.

Another example of the effectiveness of the proposed technology is the fabrication of a low-propagation-loss plasmonic waveguide made from a “plasmonic-unfriendly” $Co$ and the integration of the $Co$- based plasmonic waveguide with a $Si$ nanowire waveguide.  The cobalt and iron are ferromagnetic metals, which have large magneto-optical coefficients. The use of a ferromagnetic metal in a plasmonic structure is essential for the fabrication of an effective plasmonic isolator \cite{Shimizu2018,Shimizu2022,Zayets2012Isolator,Zayets2015,Kaihara2015,Armelles2013}.

The challenge for the integration of the low-propagation loss plasmonic waveguides with a $Si$ nanowire waveguides is that the $Si$ nanowire waveguides are fabricated on a thick $SiO_2$ layer having a very small refractive index. As is shown in \textbf{Supplement}, the reduction of propagation loss of a surface plasmon can be achieved only in a structure, in which the refractive index of a thin insertion layer is smaller than that of the dielectric substrate. It is difficult to find a material whose refractive index is substantially smaller than that of $SiO_2$. The solution to this problem is the etching out and the replacement of $SiO_2$ by a material of a higher refractive index in the plasmonic section.

\begin{figure}[h]
\begin{center}
\includegraphics[width=8cm]{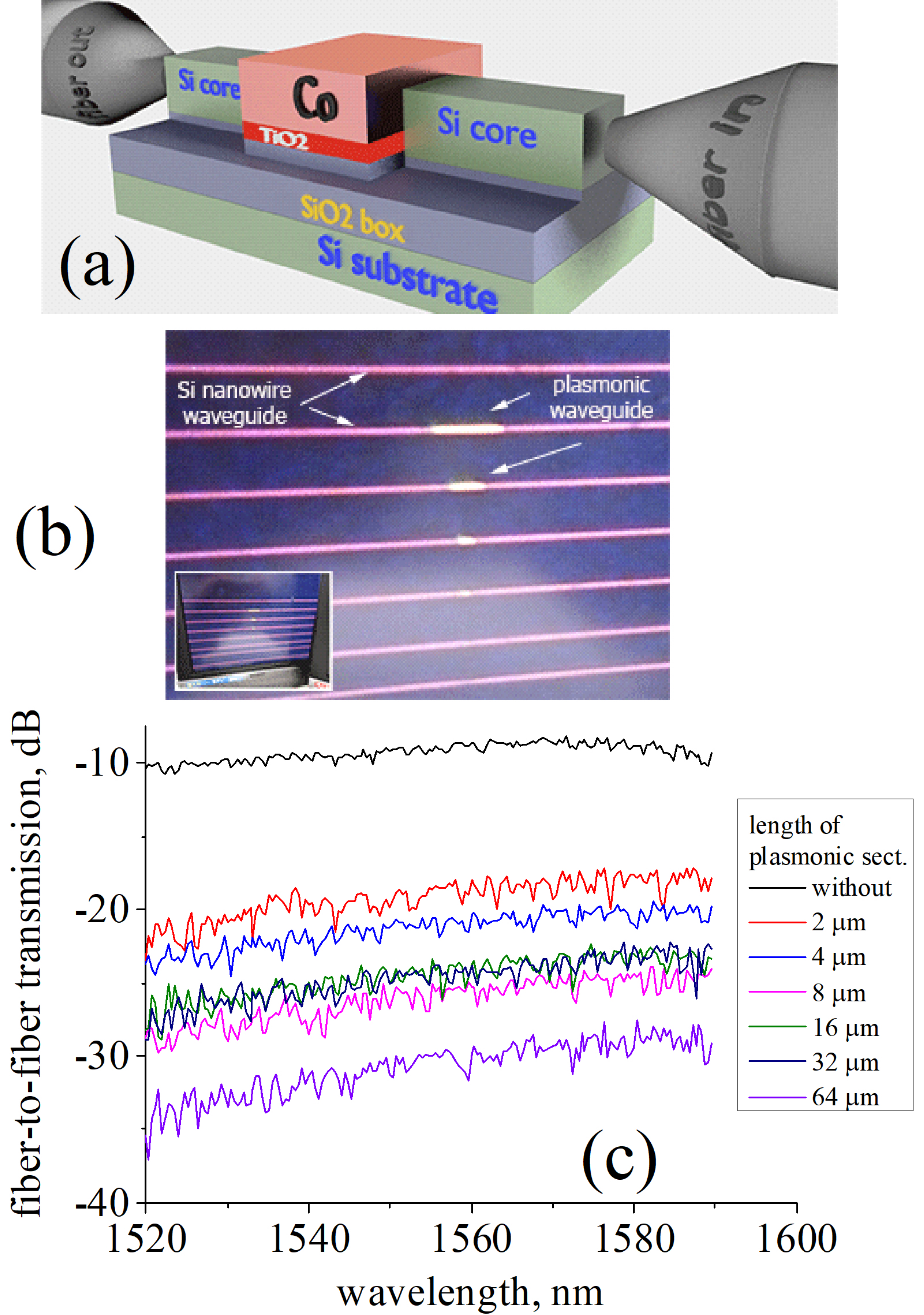}
\end{center}
\caption{\label{fig:fig5} 
(a) Experimental setup for measuring the propagation loss of a surface plasmon; (b) Top- view image of the fabricated waveguides. Each waveguide has a different length of the plasmonic section; (c) Fiber-to-fiber transmission as a function of wavelength measured for a different length of the plasmonic section \cite{Shimizu2018}.  
}
\end{figure}


 We have used $TiO_2$ as a material of a high refractive index and $SiO_2$ as a material of a thin insertion layer. Details of the fabricated structure and the fabrication technology are described in \textbf{Supplement}. Figure \ref{fig:fig5}(a) shows the measurement setup for evaluation of the plasmon' propagation loss. The fabricated device consists of a $Si$ nanowire waveguide, a part of which is etched out, and the consequent layers of  $TiO_2$, $SiO_2$  and $Co$ are deposited into the gap. A set of identical waveguides having a different length of the plasmonic section were fabricated on the same wafer. Figure \ref{fig:fig5}(b) shows a top view of the fabricated device. Figure \ref{fig:fig5}(c) shows the measured fiber-to-fiber transmission as a function of the wavelength for different lengths of the plasmonic waveguide.  From this dependence the propagation loss of a surface plasmon is evaluated to be 0.7 dB/$\mu m$.  Extrapolation of the propagation loss to the zero length of the plasmonic section gives the coupling loss between the plasmonic and the Si nanowire waveguides of 4 dB per a facet \cite{Shimizu2018}.
 
The measured propagation loss is still above the theoretical limit of 0.01 dB/$\mu m$, but is already sufficiently small for a practical application. For example, the achieved propagation loss in a plasmonic structure containing "plasmon-unfriendly" cobalt or iron is smaller than that in conventional gold/dielectric plasmonic structure. (See Fig.\ref{fig:fig1}(b)). 

The propagation loss can be further reduced when the Atomic Layer Deposition (ALD) is used instead of sputtering for deposition of the thin insertion layer.  ALD allows for extremely precise control of the film thickness and uniformity, which are both critically important for the reduction of the propagation loss of a surface plasmon.


In conclusion, the fabrication technology of a low- loss plasmonic waveguide is described. The proposed technology allows the use of different metals having different unique properties in low-loss plasmonic structures, additionally to traditionally-used gold.

The physical principle of the loss reduction is the redistribution of the optical field of a surface plasmon across the metal/dielectric interface.  A reduction of the optical confinement and an increase of the volume of surface plasmon lead to a decrease of the optical field inside the metal and, as a result, to a decrease of the propagation loss. An optimization of the optical confinement can be achieved either by an insertion of a thin dielectric layer\cite{Zayets2012Isolator,2011SiSiOCu} or by the modulation of interface roughness\cite{Shimizu2018} or sharpness or by an optimization of the in-plane confinement of a surface plasmon\cite{2011SiSiOCu,Shimizu2018}.

The effectiveness of the proposed technology was experimentally demonstrated for a loss reduction in a plasmonic structure containing   “plasmon- unfriendly” cobalt.  The plasmon’s propagation loss was reduced from 11 dB/$\mu m$ down to 0.7 dB/$\mu m$ in a Co-base plasmonic structure. 

\bibliography{Plasmon}

\section{Supplementary Material}

\textit{\underline{Abstract}} The supplement describes examples of a reduction of the propagation loss of a surface plasmon in a different plasmonic structure. It demonstrates that a substantial reduction of the propagation loss of a surface plasmon is a general rule for a variety of a different material of the metal and the dielectric. The propagation constants and the field distribution of a surface plasmon is calculated for a \mbox{$Co/SiO_2/TiO_2$}  plasmonic structure. Similar to a \mbox{$Co/SiO_2/Si$}  plasmonic structure, which is described in the main part of the  manuscript, the \mbox{$Co/SiO_2/TiO_2$} plasmonic structure has an optimum $SiO_2$ thickness, at which the plasmonic propagation loss is reduced below 0.01 dB/$\mu m$. In both cases, the refractive index of a thin  inserted layer of $SiO_2$ is smaller than that of the dielectric ($TiO_2$ or $Si$). In contrast,  the insertion of a thin layer of a larger refractive at the metal/dielectric interface does not lead to a substantial reduction of the plasmonic loss. As an example,  a \mbox{$Co/TiO_2/SiO_2$} plasmonic structure is studied, in which the propagation loss of a surface plasmon is calculated to be substantial independently of the  $TiO_2$ thickness. It is because the refractive index of the inserted  $TiO_2$ thin layer is larger than that of $SiO_2$. Additionally, a \mbox{$Co/SiO_2/TiO_2/SiO_2/Si$} plasmonic structure was studied and optimized. This plasmonic structure was used in our experimental verification  of the proposed method of the loss reduction.  The details of the fabrication technology of the integration of a Co-based plasmonic waveguide and a Si nanowire waveguide were explained.

\begin{figure}[htbp]
\begin{center}
\includegraphics[width=8.2cm]{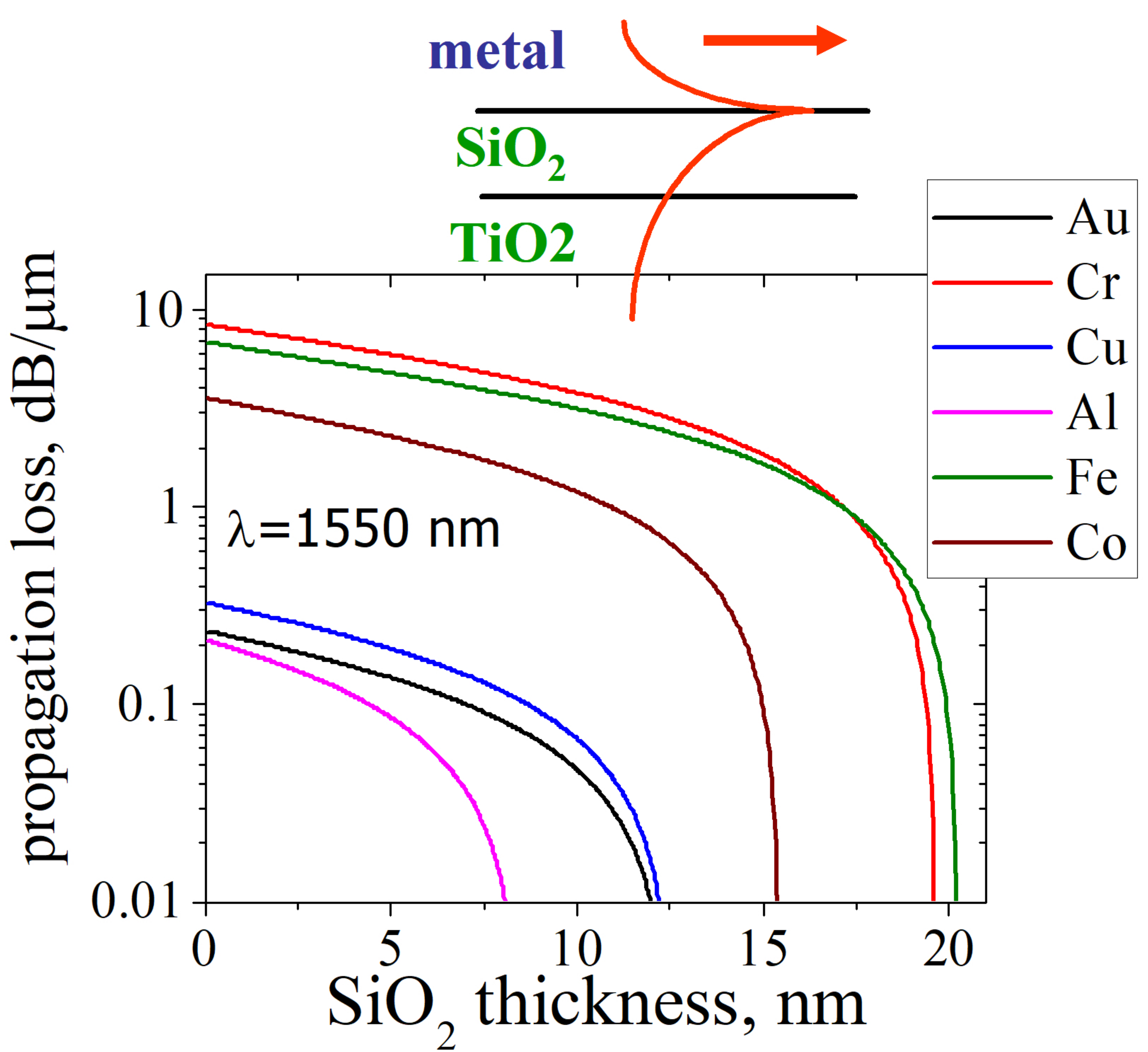}
\end{center}
\caption{\label{fig:figS1} 
Propagation loss of a surface plasmon propagating along a $metal/SiO_2/TiO_2$ interface. For each metal there is an optimum $SiO_2$ thickness, at which plasmonic loss becomes negligibly small.  $\lambda$ =1550 nm
}
\end{figure}


\subsection{Insertion layer of a smaller refractive index effective for reduction of  plasmonic loss}

In the case when the refractive index of a thin insertion layer at the metal/dielectric interface is smaller than the refractive index of the dielectric, the propagation loss of a surface plasmon can be reduced below 0.01 dB/$\mu m$. The reason for the reduction is that the insertion layer redistributes the optical field of the surface plasmon near the  interface pushing it out of the metal and deeper into the dielectric. 

The \mbox{$metal/SiO_2/TiO_2$} plasmonic structure is an example of such a structure. The refractive index of $SiO_2$  ($n_{SiO2}$=1.444  at  $\lambda$ = 1550 nm) \cite{nSiO2} is smaller than the refractive index of  $TiO_2$ ($n_{TiO2}$=2.4538 at $\lambda$ = 1550 nm) \cite{nTiO2}. Figure \ref{fig:figS1} shows the calculated propagation loss in the \mbox{$metal/SiO_2/TiO_2$} plasmonic structure as a function of the $SiO_2$ thickness. Similar to the \mbox{$metal/SiO_2/Si$} plasmonic structure (See Fig.3), there is an optimum $SiO_2$ thickness, which is different for each metal and  at which the plasmonic propagation loss can be reduced below 0.01 dB/$\mu m$.  For the same metal, the optimum $SiO_2$ thickness is larger for the \mbox{$metal/SiO_2/TiO_2$} plasmonic structure than for the \mbox{$metal/SiO_2/Si$} plasmonic structure. The slopes of curves near the optimum $SiO_2$ thickness  are smaller in the case of the \mbox{$metal/SiO_2/TiO_2$} structure. It means that the requirements for the roughness of $SiO_2$ layer are softer for the \mbox{$metal/SiO_2/TiO_2$} structure.

\begin{figure}[htbp]
\begin{center}
\includegraphics[width=8.2cm]{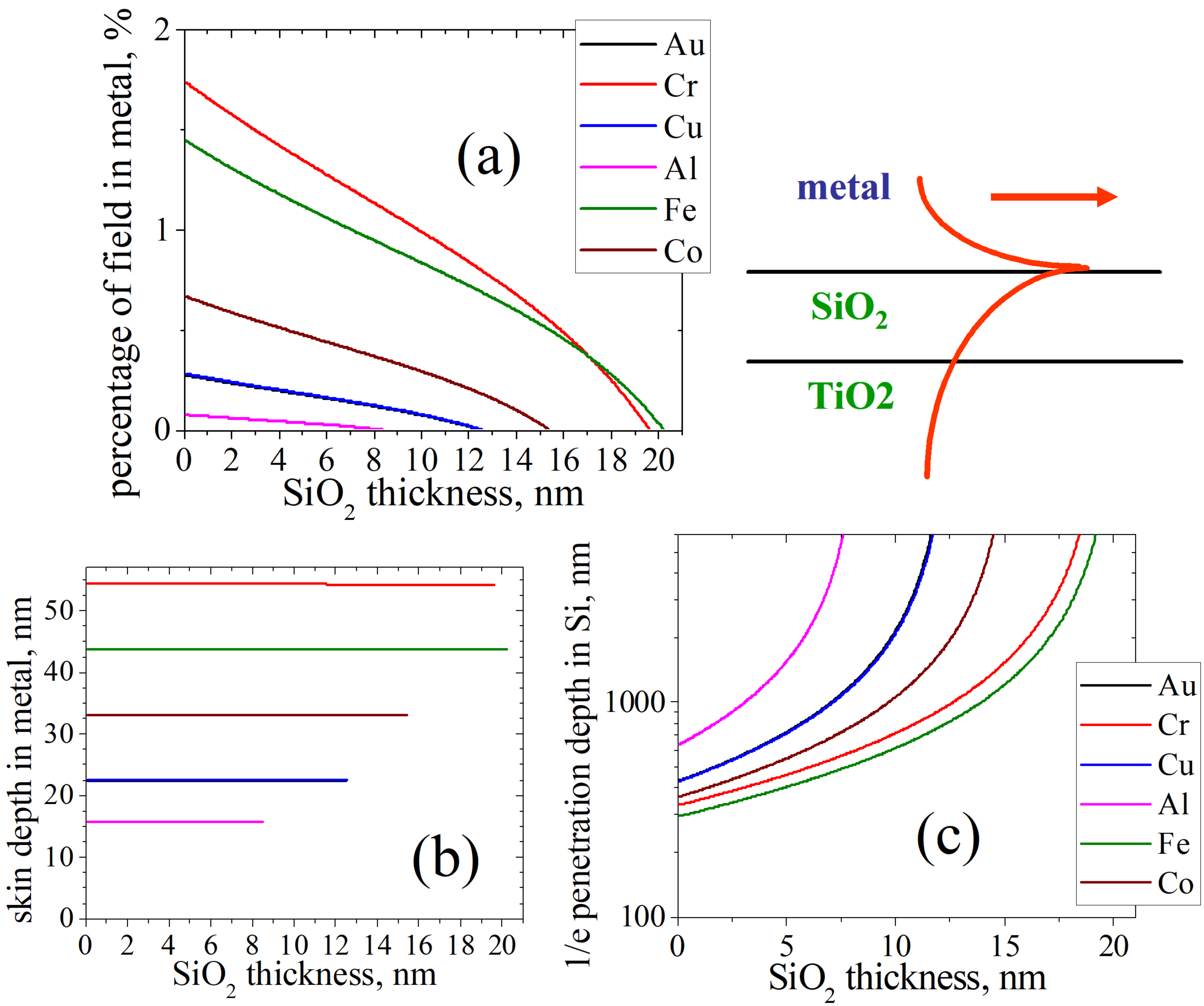}
\end{center}
\caption{\label{fig:figS2} 
(a) Surface plasmon at a $metal/SiO_2/TiO_2$ interface. (b) Percentage of optical field of the plasmon, which is inside of the metal; (b) skin depth in the metal; (c) 1/e penetration depth in $TiO_2$. $\lambda$ =1550 nm.
}
\end{figure}


Figure \ref{fig:figS2}(a) shows the calculated  percentage of light penetrating into the metal as a function of the $SiO_2$ thickness. The dependency is very similar to that of Fig.4(a). At the optimum $SiO_2$ thickness the penetration is the smallest. As a result,  there is a very small amount of optical field inside the metal and the propagation loss of a surface plasmon is small.
 
Figures \ref{fig:figS2}(b) and Figure \ref{fig:figS2}(c) show the skin depth in the metal and the penetration depth in the dielectric. Similar to Fig.4(b), the skin depth in the metal is practically independent of the $SiO_2$ thickness, but the penetration depth in the dielectric increases substantially near the optimum $SiO_2$ thickness. From compression of Figs. 4(b) and \ref{fig:figS2}(b), the skin depth changes only slightly between two plasmonic structures and practically can be considered as independent of the dielectric material. 

\begin{figure}[htbp]
\begin{center}
\includegraphics[width=8.2cm]{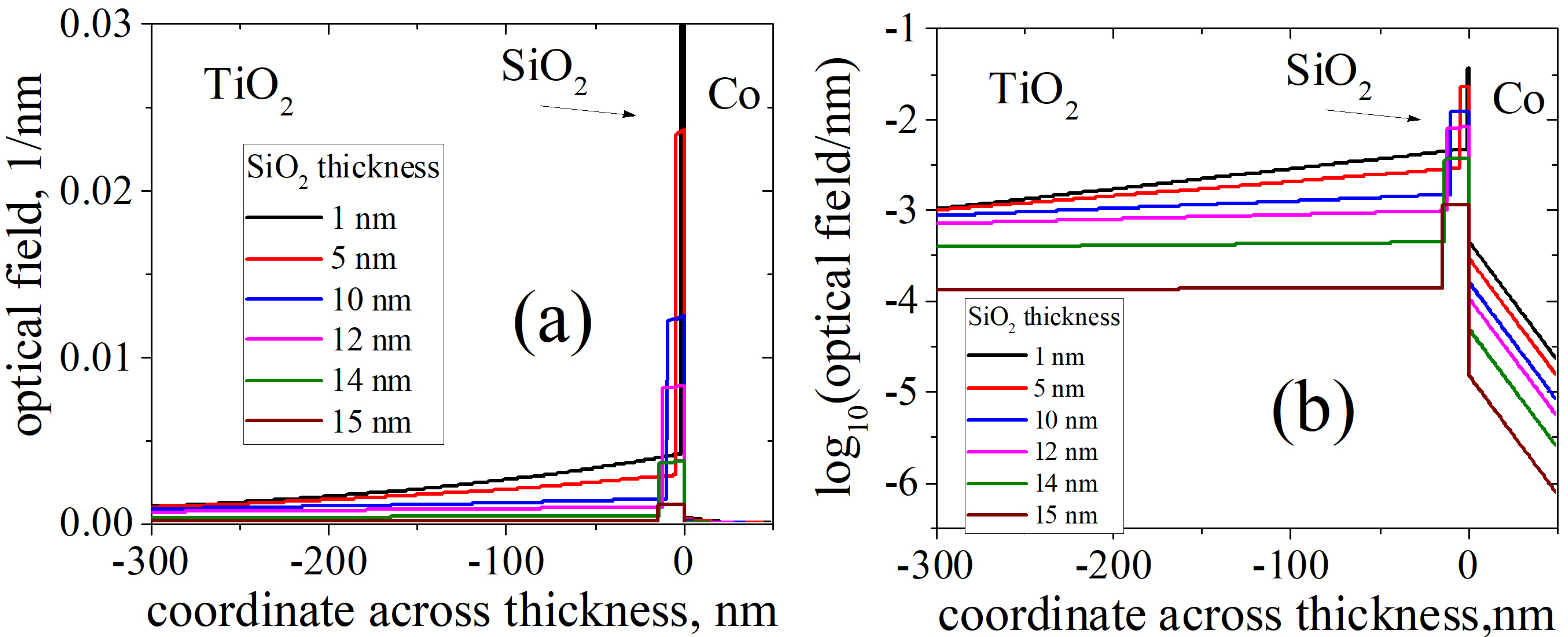}
\end{center}
\caption{\label{fig:figS3} 
Distribution of the optical field across thickness in a \mbox{$Co/SiO_2/TiO_2$} plasmonic structure at a different $SiO_2$ thickness (a) linear scale (b) logarithmic scale;  $\lambda$ =1550 nm.
}
\end{figure}

In both structures, the \mbox{$Co/SiO_2/TiO_2$}  and the \mbox{$Co/SiO_2/Si$}, the insertion of a thin $SiO_2$ layer causes a shift of the optical field out of the metal into the dielectric. The reason why the optical field is pushed out from the metal can be understood from Fig. \ref{fig:figS3}. The amplitude of the optical field in the inserted thin $SiO_2$ is an order of magnitude higher than in $TiO_2$. As a result, even though the insertion layer is very thin, it substantially affects the propagation constants of a surface plasmon. The effective refractive index $n_{plasmon}$ of a surface plasmon is slightly higher than the refractive index of $TiO_2$. The insertion of a layer of a smaller refractive index reduces $n_{plasmon}$. When the $SiO_2$ thickness increases, there is more optical field inside $SiO_2$ and  $n_{plasmon}$  becomes smaller and, therefore, more close to the refractive index of $TiO_2$. As a consequence, the  perpendicular- to - interface component of the wave vector in $TiO_2$ becomes smaller and, therefore, the penetration depth in $TiO_2$ increases. At the cut-off thickness,  the effective refractive index of the surface plasmon becomes equal to the refractive index of $TiO_2$ and the surface plasmon loses its optical confinement to the metal surface.

\subsection{Insertion layer of a larger refractive index ineffective for loss reduction }

\begin{figure}[htbp]
\begin{center}
\includegraphics[width=8.2cm]{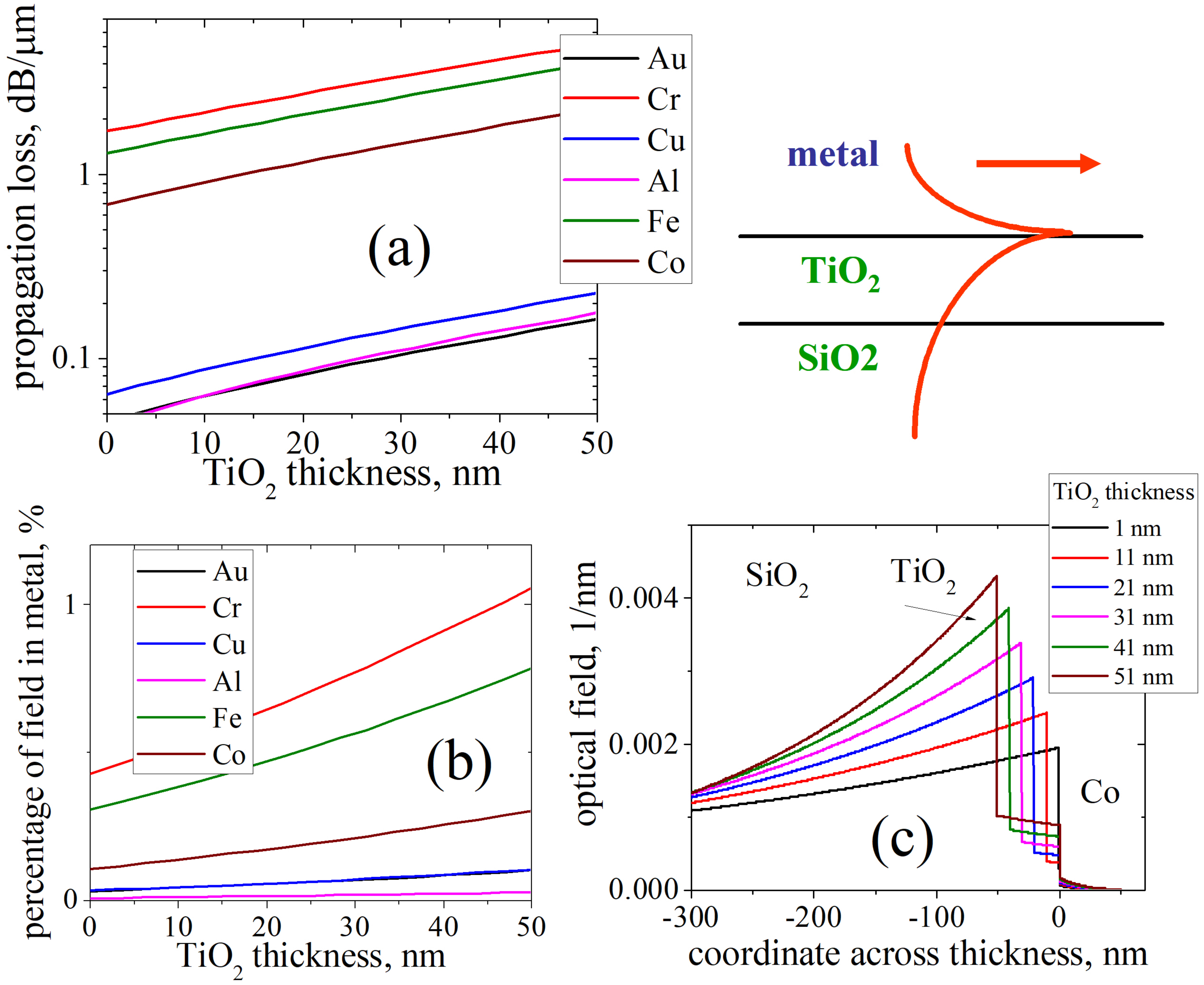}
\end{center}
\caption{\label{fig:figS4} 
 Surface plasmon at a $Co/TiO_2/SiO_2$ interface. (a) Propagation loss of a surface plasmon. (b) Percentage of optical field of the plasmon, which is inside of the metal; (c) Distribution of optical field across thickness  at a different $TiO_2$ thickness. $\lambda$ =1550 nm.
}
\end{figure}

A thin insertion layer at a metal/dielectric interface, whose refractive index is larger than the refractive index of the dielectric, is not effective for the reduction of propagation loss of a surface plasmon. A $Co/TiO_2/SiO_2$ plasmonic structure is studied as an example of such an ineffective plasmonic structure. Figures \ref{fig:figS4}(a) shows the calculated propagation loss as a function of $TiO_2$ thickness. The insertion of  the $TiO_2$ thin layer does not reduce the propagation loss. The tendency is opposite to that of  Fig. \ref{fig:figS1} and Fig.4 as the propagation loss increases when the $TiO_2$ thickness increases. The loss increases, because the percentage of the optical field in the metal increases (See Fig. \ref{fig:figS4}(b)).

The reason why the insertion of a thin $TiO_2$ does not lead to reduction of the propagation loss of a surface plasmon is the following. The effective refractive index $n_{plasmon}$ of a surface plasmon in the $Co/SiO_2$ plasmonic structure is slightly larger than the refractive index $n_{SiO2}$  of $SiO_2$. The 1/e penetration depth of the evanescent field in the $SiO_2$ layer is larger, when the difference between $n_{plasmon}$ and $n_{SiO2}$ is smaller. The insertion of a thin $TiO_2$ layer of a higher refractive index makes the plasmon' propagation constant $n_{plasmon}$ larger and, therefore, the difference between  $n_{plasmon}$ and $n_{SiO2}$ becomes even larger. As a result, the 1/e penetration depth into $SiO_2$ decreases, which leads to a larger amount of optical field inside the metal and, therefore, to a larger propagation loss of a surface plasmon. Figure \ref{fig:figS4}(c) shows the distribution of the optical field across the interface for the $Co/TiO_2/SiO_2$ plasmonic structure.

\subsection{Technological considerations for integration of plasmonic and silicon nanowire waveguides}

In order to be competitive, the plasmonic devices must be integrated into the Photonic Integrated Circuits (PIC) and, therefore, the fabrication technology of a plasmonic device should be compatible with the existing fabrication technology of PIC. The standard fabrication technology of PIC uses a Si(220 nm)/$SiO_2$(3000 nm)/Si wafer (SOI wafer) and a plasmonic device should be fabricated on the same SOI wafer.  The problem of the SOI wafer for the application of the proposed reduction method of the plasmonic loss is very small refractive index of $SiO_2$ ($n_{SiO2}$=1.444  at  $\lambda$ = 1550 nm) \cite{nSiO2}. As was discussed above, the refractive index of the insertion layer should be smaller than the refractive index of the dielectric layer. It is difficult to find a material whose refractive index is substantially smaller than that of $SiO_2$. The solution of this problem is the etching out and the replacement of $SiO_2$ by a material of a higher refractive index in the plasmonic section.

The first option for the material of a high refractive index, which we have tested, was silicon. The refractive index of $Si$ is high ($n_{Si}$=3.477 at $\lambda$ = 1550 nm) \cite{nSi}. Following etching of $SiO_2$ by the Inductively Coupled Plasma - Reactive Ion Etching (ICP-RIE ), $Si$ and a thin $SiO_2$ insertion layer  (See Fig.3) were deposited. However, the measured plasmon' propagation loss was high of about 4 dB/$\mu m$.  Presumably, the high loss is due to a substantial bulk optical absorption of the sputtered $Si$.

\begin{figure}[htbp]
\begin{center}
\includegraphics[width=8.2cm]{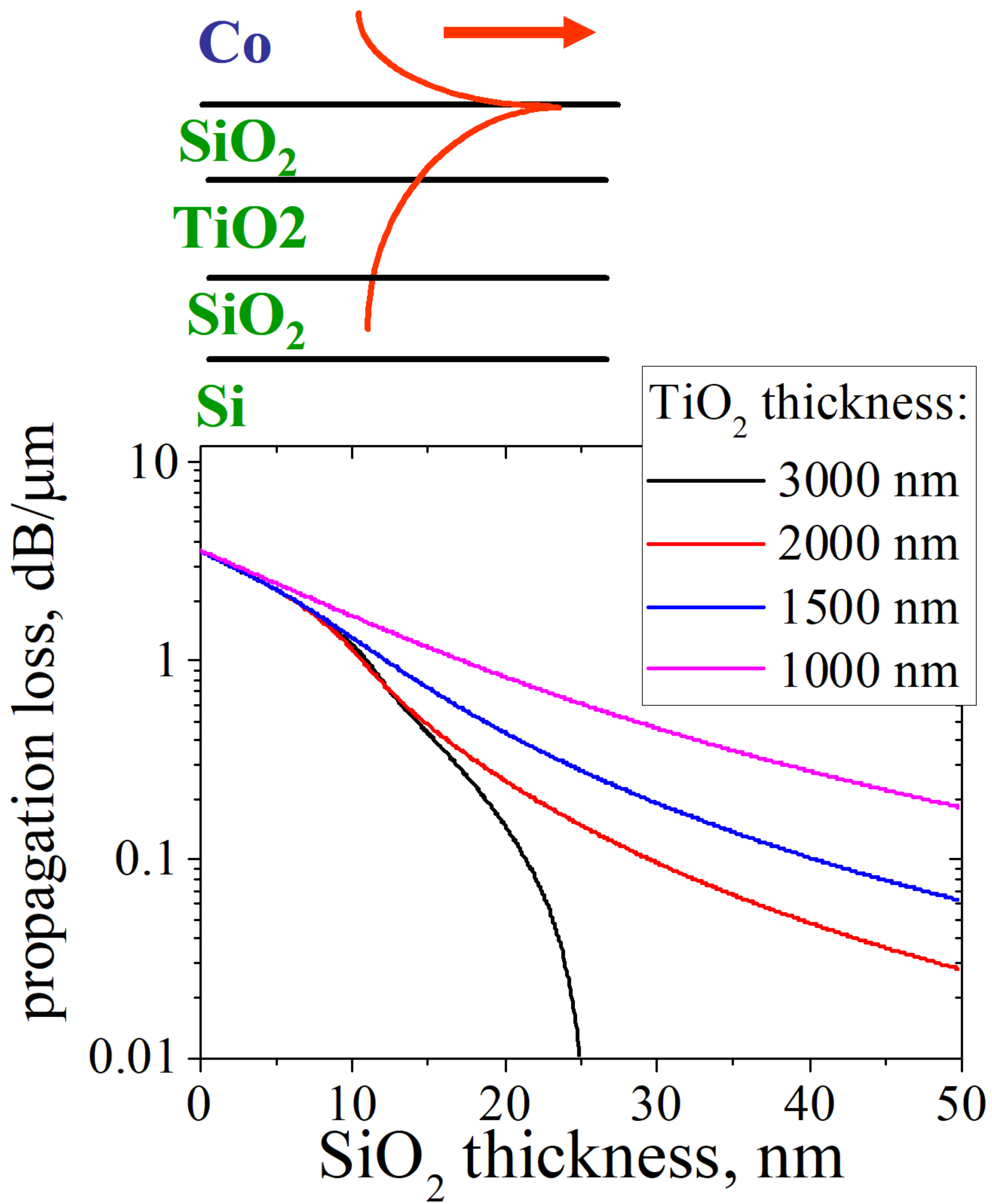}
\end{center}
\caption{\label{fig:figS5} 
Propagation loss of a surface plasmon in a $Co/SiO_2/ TiO_2/SiO_2/Si$ structure as a function of thickness of the top $SiO_2$ layer. Each curve corresponds a different thickness of $TiO_2$ when the sum of thicknesses of the  $TiO_2$ and bottom $SiO_2$ layers is kept to be 3 $\mu m$.  $\lambda$ =1550 nm
}
\end{figure}

The second option for the material of a high refractive index, which we have tested, is $TiO_2$. The refractive index of $TiO_2$ is smaller than that of $Si$ , but still high ($n_{TiO2}$=2.4538 at $\lambda$ = 1550 nm) \cite{nTiO2}. Even though the achieved material quality of the sputtered $TiO_2$ is better that that of $Si$, it is still technologically difficult to obtain the required material quality of $TiO_2$ in a deep-etched micro-sized plasmonic section.  A reduction of the etching depth and the $TiO_2$ thickness are desirable to keep a required interface quality. Figure \ref{fig:figS5} shows the calculated propagation loss of a surface plasmon for a different etch depth and, correspondingly, a different thickness of $TiO_2$. The black line shows the case when the 3-$\mu m$-thick $SiO_2$ is fully etched. In this case, the propagation loss can be below 0.01 db/$\mu m$, when thickness of the top  $SiO_2$ is about 25 nm. Even in the case when only a half of the $SiO_2$ layer is etched, the propagation loss is reasonably low.

\subsection{Details of  fabricated structure and fabricated technology}

The device shown in Fig.5 was fabricated on a SOI wafer. A sputtered 40-nm-thick $SiO_2$ was used as a hard mask. Following an electron-beam (EB) lithography, a 450-nm-wide Si nanowire waveguide with   a 150-nm-wide spot-size converter was dry etched by an ICP-RIE. Next, a 2 $\mu m$ of $SiO2$ was dry-etched by an ICP-RIE in the plasmonic section and a  $TiO2$ (2.1 $\mu m$) / $SiO2$ (40 nm) / $Si$ (100 nm)  was sputtered and lift-off, so that it remains only in the plasmonic section. Next, following the EB lithography, a 70-nm-wide Si bridge was etched in the plasmonic section. The Si bridge is required for the in-plane confinement of a surface plasmon \cite{Shimizu2018,ZayetsSPIE2022}. Finally,  $SiO2$ (10 nm) / $Co$ (200 nm)/$Au$ (500 nm) was sputtered into the plasmonic section.



\end{document}